\documentstyle[aps,twocolumn,epsfig,amsmath]{revtex}

\input epsf
\epsfverbosetrue

\draft

\begin{document}

\title{Geometrical Dependence of High-Bias Current in Multiwalled Carbon Nanotubes}

\author{B. Bourlon,$^1$ D.C. Glattli,$^{1,2}$ B. Pla\c{c}ais,$^1$ J.M. Berroir,$^1$ L. Forr\'{o},$^3$ A. Bachtold$^{1*}$}

\address{
$^1$ LPMC, Ecole Normale Sup\'{e}rieure, 24 rue Lhomond, 75005 Paris,
France. $^2$ SPEC, CEA Saclay, F-91191 Gif-sur-Yvette, France.
$^3$ EPFL, CH-1015, Lausanne, Switzerland. }

\date{ \today}
\maketitle

\begin{abstract}

We have studied the high-bias transport properties of the
different shells that constitute a multiwalled carbon nanotube.
The current is shown to be reduced as the shell diameter is
decreased or the length is increased. We assign this geometrical
dependence to the competition between electron-phonon scattering
process and Zener tunneling.

\end{abstract}

\vspace{.3cm} 

Building electronic devices at the molecular scale has motivated
intense research for the last years. Carbon nanotubes have been
used as building blocks for devices such as nanotube-nanotube
junctions \cite{fuhrer,yao1} or field-effect transistors
\cite{tans,martel} that have been integrated in logic circuits
\cite{bachtold1,derycke,liu}. Recently nanotubes have also
attracted widespread attention as future
interconnects\cite{mceuen,wei,kreupl,li}, since their
current-carrying capacities are several orders of magnitude larger
than in present-day interconnects. The current in metal
single-wall nanotubes (SWNT) saturates at $20-25$ $\mu$A, which
corresponds to a current density exceeding $10^{9}$ A/cm$^{2}$
\cite{yao2}.

An important question that remains to be addressed is how the
current saturation grows with the section of an interconnect made
of nanotubes. The interconnect may be made larger by selecting
SWNTs assembled in a bundle. However, the intertube transmission
is weak and leads to a bad current distribution over the whole
bundle when the current is injected in one tube. Another approach
is to choose multiwalled nanotubes (MWNTs) with many concentric
shells. The intertube transmission within a single MWNT is larger.
The interface of two neighbor shells is maximum and equal to the
entire shell surface. Recently, it has been shown that most of the
shells contribute to the current at large bias \cite{bachtold2}
and that the current of each shell saturates like in SWNTs
\cite{collins1,collins2}. Remarkably, the saturation value was
reported to be close to what is obtained for SWNTs and to be
independent on the shell diameter. This is very surprising since
more subbands are expected to carry current in large shells. MWNT
shells can be much larger than SWNTs, reducing substantially the
energy separation between subbands.

In this Letter, we present studies of the high-bias transport in
MWNTs with different electrode separations $L$. We find that the
current saturation per shell varies between $10$ and $60\mu$A. For
the shortest separation ($200$ nm), the current is shown to
increase as a function of diameter. For longer separations
($1\mu$m), the diameter dependence is much weaker and more
difficult to observe. We propose that this geometrical dependence
is due to a weak variation of the number of current carrying
subbands within each shell that results from the competition in
the electron transmission between the electron-phonon scattering
process and Zener tunneling. The Landauer formula is applied that
incorporates these two processes and the numerical calculations
reproduce well the experimental results. Interestingly, the model
describes the experimental observation that the high-bias current
does not depend on the metal or the semiconducting character of
the shell at large diameters.

The MWNTs were synthesized by arc-discharge evaporation and
carefully purified \cite{bonard}. The MWNTs are sonicated in
dichloroethane and dispersed onto an oxidized Si substrate. Atomic
force microscopy is used to image, select and locate nanotubes
that are then contacted with Cr/Au electrodes by electronic
lithography. An example of a device is shown in Fig 1(a). Typical
two-point resistances at low-bias vary in the range $5$ to
\mbox{$30$ k$\Omega$}.

The electronic properties of different shells in a MWNT are probed
using the electrical-breakdown method developed by Collins $et$
$al.$ \cite{collins1,collins2}. The shells are selectively removed
stepwise by injecting a large current in the MWNT. The bias
voltage is increased until a sharp step in the current is detected
that corresponds to the failure of one shell. Then, the bias
voltage is quickly put down to 0 in $100$ ms. The removed shell
has been shown to be the outermost one which is in direct contact
with air and therefore most subject to a rapid oxidation initiated
by the current \cite{collins2}. We remove shells until no more
current is detected. The final diameter has then decreased to
typically $2-4$ nm as seen on the AFM picture in Fig. 1(b). The
image shows that the peeling is uniform along the tube, except
near the electrodes. Note that the failure is apparently different
for the last shells since they are not evaporated. Failure may be
local and situated at a small gap, too short to be observed with
AFM.

\begin{figure}[bht]
\epsfxsize=80mm%
\centerline{\epsfbox{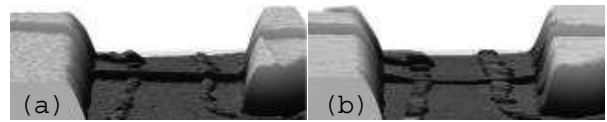}}
\vspace{2mm}
 \caption{
AFM image of a MWNT (a) before and (b) after the application of
the electrical-breakdown method. The diameter has been reduced
from $12$ to $3$ nm. The electrode separation is $600$ nm and the
electrode height $45$ nm.
 }
\end{figure}

We have designed an experiment where a series of electrodes is
attached to a single long MWNT with various separations from $200$
to $1200$ nm.  Fig. 2(a-c) show the result of the
electrical-breakdown method applied to the different sections of
this MWNT. Let us first consider the current-voltage
characteristics ($I-V$) of the pristine sample given by the upper
$I-V$ curve. It appears that the high-bias current is
significantly larger in the shortest section. This result can be
enlightened by analyzing the current carried by the different
shells of the MWNT that are obtained by comparing the successive
$I-V$s. Inspection of Fig. 2(a-c) shows that the current increase
in the 200 nm section is the combined effect of $(i)$ more jumps
and $(ii)$ larger jumps.

\begin{figure}[bht]
\epsfxsize=80mm%
\centerline{\epsfbox{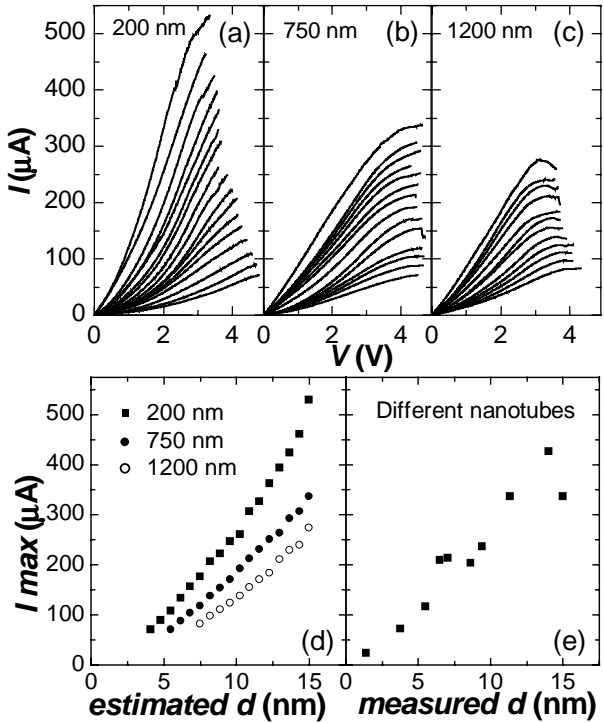}} \vspace{2mm}
 \caption{
(a-c) High-bias $I-V$ characteristics of a same $15$ nm diameter
MWNT with different electrode separations. Each $I-V$ corresponds
to the loss of one shell. The 11th $I-V$ in (b) and the 5th $I-V$
in (c) could not be recorded. The $I-V$s have been measured at
room temperature in air. (d) Current maximum as a function of the
estimated diameter from (a-c). (e) Current maximum as a function
of diameter for different MWNTs. The length is between $500$ and
$750$ nm.
 }
\end{figure}

We discuss first the jump number. This 15 nm diameter MWNT is
expected to consist of $\sim 19$ shells. The jump number is
however smaller and $L$ dependent. It is reduced from 17 to 13
when $L$ passes from 200 to 1200 nm. A smaller number of jumps
might be due to a smaller number of shells and to a thinner
diameter. However, this is unlikely since AFM imaging indicates a
uniform diameter over the entire length. This observation suggests
rather that some shells do not participate to the current,
especially in the longer sections. Such an interpretation is
supported by the following observation in Fig. 2(d), where the
maximum current of each $I-V$ is plotted as a function of the
diameter $d$. The diameter has not been measured by AFM at each
step but estimated assuming that the diameter decreases each time
by twice the interlayer distance $3.4$ \AA. The extrapolated
curves should pass by \mbox{$25$ $\mu$A} at $1.4$ nm, the
saturation of a SWNT. This is not the case. Rather, the
extrapolations have negative current values at $1.4$ nm, which
indicates that $d$ is overestimated and suggests that some shells
do not participate to the current.

As mentioned before, the current jumps for the shortest length are
larger in amplitude. The jump averages are 28.7, 19.1 and 17.4
$\mu$A for respectively the 200, 750 and 1200 nm lengths.
Remarkably, the curves in Fig. 2d deviate from a linear behavior.
The deviation is best seen for the \mbox{$200$ nm} length. It
shows that large-diameter shells contribute more to current. This
finding has been observed in more than $20$ MWNTs. Other examples
are shown in Fig. 3(a) and (b), where the current jump $\Delta I$
is plotted as a function of $d$. Both curves show that $\Delta I$
is larger for larger $d$. However, the slope is much weaker for
the $650$ nm length and is hardly observed due the measurement
noise.

\begin{figure}[bht]
\epsfxsize=80mm%
\centerline{\epsfbox{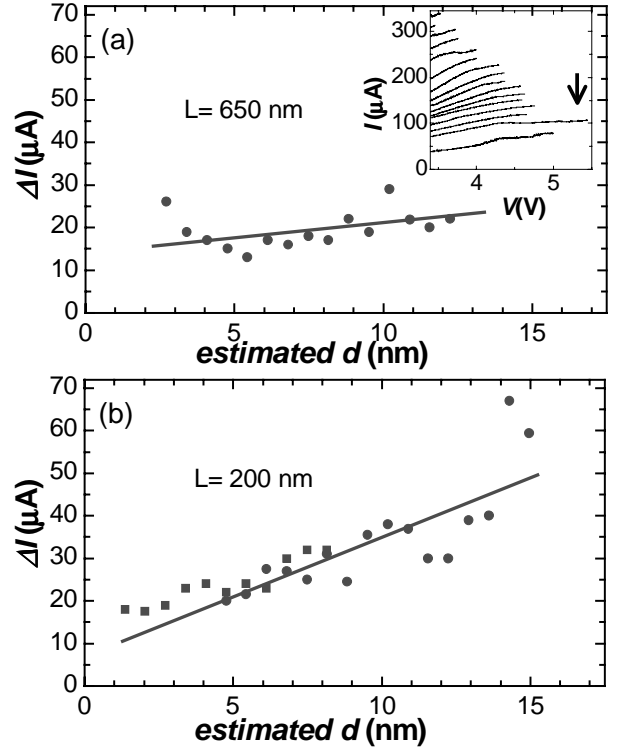}}
\vspace{2mm}
 \caption{
High-bias jump of the current as a function of diameter for (a)
$650$ nm and (b) $200$ nm lengths. The straight lines are guides
for the eyes. In (b) the data of two samples are shown. The inset
to (a) shows the $I-V$s.
 }
\end{figure}

We note that an accurate determination of the current jump is
difficult. Inset of Fig. 3(a) shows that the bias at which the
shell fails increases irregularly and that the $I-V$s are not
smooth. We define the current-jump as the current difference at a
bias $V_0$ given by the maximum voltage recorded for the two
successive $I-V$s. Another approach could have been to take the
maximum of the difference or the difference of the maximums.
However, the observation of a current increase with diameter is
not altered by the choice of the estimation procedure.
Importantly, the current jump estimation is precise enough to
obtain close slopes of $\Delta I(d)$ for different MWNTs with same
length as seen in Fig. 3(b).

We now discuss the possible origins of these geometrical
dependences in the shell current. We review first the
electron-phonon scattering process that causes the current to
saturate at $20-25 \mu$A in metal SWNTs \cite{yao2}. An electron
that is accelarated by an electric field $\mathcal{E}$ will gain
energy until an optical phonon of energy $\hbar\Omega$ is emitted
(\mbox{$\hbar\Omega\approx$ 0.16eV}). At this point, the coupling
to the phonon is so strong that the electron is backscattered.
Using a Landauer-type argument the current saturation is shown to
be proportional to the phonon energy,
\mbox{$(4e/h)\hbar\Omega\approx25 \mu$A}. The mean free path
$l_{ph}$ is equal to the distance an electron must travel to be
accelerated to $\hbar\Omega$, $l_{ph}=\hbar\Omega/e\mathcal{E}$
\cite{yao2}. The total transmission trough the whole tube is
obtained by dividing $l_{ph}$ by $L$
\begin{equation}
T_{ph}=\hbar\Omega/eV \label{phonon}
\end{equation}
where V is the voltage applied between the electrodes. Eq. (2) is
valid for large $V$.

At first sight, the current jump variations might be explained
from the energy variation of phonons as a function of diameter.
However, $ab$ $initio$ calculations have shown that optical phonon
energy changes barely with tube diameter \cite{dubay}. An
alternative mechanism is thus needed to account for the
current-jump variation.

We propose that the current sensitivity to diameter results from
the contribution of Zener tunneling \cite{anantram}. Fig 4(a)
shows a scheme that represents the potential variation in space of
the band structure. The potential drop should be in principle
determined by the self-consistent solution of the Poisson equation
and the nonequilibrium electron density. We assume here that the
voltage drop is linear, a good approximation when multiple
electron-phonon scattering processes occur \cite{rem-highRc}.

An electron that enters a non-crossing subband from the left
electrode reaches at some point the top of the valence band and
tunnels to the conduction band. The Zener-tunneling transmission
\cite{sze} is given by
\begin{equation}
T_{Z}=\exp(-\frac{4\sqrt{2m^*}LE^{3/2}}{3e\hbar V}) \label{zener}
\end{equation}
where we have assumed a linear voltage drop. $m^*$ is the
effective mass and $E$ the gap that goes like $1/d$.

Now, we compare the two transmissions as a function of diameter.
The phonon transmission is independent on $d$ and $T_{ph}=0.045$
at $3.5$ V. For a large semiconducting shell of $15$ nm diameter
and $L=200$ nm, we obtain $T_{Z}\approx0.7$ for tunneling between
the upper valence band and the lower conduction band. The
effective mass has been estimated using the thigh-binding model
for a zig-zag tube. The comparison between $T_{ph}$ and $T_{Z}$
shows that the current is limited by phonons. Zener reflexion is
here unimportant and no difference in current is expected between
metal and semiconducting shells in agreement with experiment. The
situation changes for small-diameter shells. For $d=3$ nm we
estimate $T_{Z}\approx10^{-4}$ so that the current is now limited
by Zener tunneling \cite{rem-mass}.

\begin{figure}[bht]
\epsfxsize=80mm%
\centerline{\epsfbox{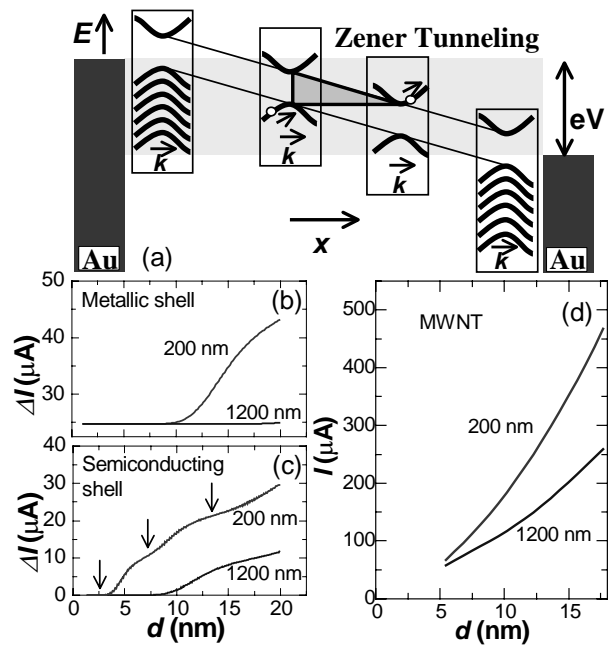}}
\vspace{2mm}

 \caption{
(a) Schematic of the potential variation in space. The boxes show
the band diagram of a semiconducting shell. Numerical calculations
of the current for (b) metal and (c) semiconducting shells at
$3.5$ V and for two lengths. (d) Calculated current of a MWNT as a
function of diameter.
 }
\end{figure}

To put the above analyze on a more quantitative basis, we use the
Landauer formula to estimate the high-bias current of a shell.
Since the considered transmissions are weak, the total
transmission of the subband $i$ is given by
\mbox{$T_i^{-1}=T_{Z}^{-1}(E_i)+T_{ph}^{-1}$}. $E_i$ is the energy
separation between the $i$ valence band to the $i$ conduction
band, where $i$ is counted from the neutrality point. Note that
alternative processes such as tunneling from the $i$ valence band
to the first conduction band changes barely the results presented
below. The current is given by \mbox{$I=(2e/h)\int \sum_i T_idE$}.
The integral is calculated between $0$ and $eV$. The use of the
Landauer formula in presence of inelastic scattering in the
non-equilibrium regime is rather crude since it does not account
for the exclusion principle. But it allows to give some physical
insight. A model based on Boltzmann equation as in \cite{yao2}
that incorporates Zener tunneling is probably needed for more
accuracy.

Fig 4(b) and (c) show the calculated current as a function of
diameter of metal and semiconducting shells. The current increases
with diameter and decreases with length, in agreement with
experiment. Indeed, larger diameters decrease the subband gap and
enable better Zener transmission. Similarly, the tunneling
transmission becomes weaker for longer nanotubes as the Zener
barrier length increases.

Fig. 4(d) shows the calculated current as a function of the
diameter $D$ of MWNTs. We assume that MWNTs consist of shells that
carry current in parallel. The current through the MWNT is the sum
of the current of a series of shells with diameter ranging from
$2$ nm to $D$. For each shell an average is taken on the current
of a semiconducting and a metal shell with the statistical weights
$\frac{2}{3}-\frac{1}{3}$. The resemblance to the experiment is
remarkable. $No$ $fitting$ parameter has been used\cite{rem-iv}.

Here, we discuss the number $N$ of subbands within a shell that
carry significantly current. The calculated curve $\Delta I(d)$ in
Fig. 4c has 3 plateaus indicated by arrows. They correspond to
situations where 0, 1 and respectively 2 subbands carry current.
These subbands are close to the neutrality point and the Zener
barrier is negligible compared to phonon scattering,
$T_Z(E)>T_{ph}$. The other subbands that have no contribution to
the current correspond to deeper valence bands with large $E$.
Zener reflection has thus a dramatic role in the high bias
transport in MWNTs. About $eV/E_{sub}\approx60$ subbands in a $14$
nm shell are counted for $eV=3.5$ V and have to be taken into
account for the current estimation, but very few contribute
effectively.

Fig. 4(c) shows the current of semiconducting shells as a function
of diameter. It appears that thin diameter shells do not carry
current. This happens when $T_Z(E)<T_{ph}$. Moreover, the diameter
at which semiconducting shells start to conduct becomes larger for
longer tubes. This may explain the discussed experimental
observation that the number of jumps is reduced for longer
section. Such an interpretation is further supported as follows.
Inset of Fig. 3(a) shows that the $I-V$ indicated by an arrow
fails at a bias larger than the others. Such $I-V$s have been
observed several times but mostly for the last $I-V$. At the
failure the current is typically $50-100$ $\mu$A and corresponds
to several shells. We propose that the outermost shell is
semiconducting in this situation and do not carry current. The
current has to pass through inner shells that are protected from
oxygen and that can sustain higher voltages. At the failure, the
breakdown is mostly so violent that all the shells fail together.

We finish with a discussion on the performances of large-diameter
MWNTs as interconnects. We have shown that the number of
current-carrying subbands within a shell depends weakly on the
diameter and the length and stays typically close to one. Thus,
the current saturation of MWNTs is expected to vary roughly
linearly with the shell number and thus the diameter. This is
observed in Fig. 2(e), where the maximum current as a function of
diameter is plotted for 10 samples. The functional form is not
quadratic as it would be expected for a plain wire. This makes
current-carrying capacities of MWNTs less efficient as
interconnects when they become larger. However, the current
carrying capacity of large MWNTs still surpasses Cu wires
($\approx10^{6}$ A/cm$^{2}$). For example, the current of a $100$
nm diameter MWNT is estimated to be $10-100$ times larger.

The authors wish to thank N. Regnault, P. Morfin and C. Delalande.
The research has been supported by the CNRS, DGA, ACN programs,
Paris $6$ and Paris $7$.

$^{*}$ corresponding author: bachtold@lpmc.ens.fr


\begin{references}
\bibitem{rem-highRc}
The voltage drops are neglected at the nanotube-electrode
interface. Indeed large contact resistances are observed to have
no influence on the current saturation.
\bibitem{rem-iv}
The calculations show that the current does not really saturate
for short tubes such as $200$ nm long in contrast to the longer
ones. The calculations reproduce well the $I-V$s shown in Fig.
1(a-c).
\end{references}
\end{document}